\documentclass[fleqn,usenatbib]{mnras}

\usepackage{newtxtext,newtxmath}

\usepackage[T1]{fontenc}

\DeclareRobustCommand{\VAN}[3]{#2}
\let\VANthebibliography\thebibliography
\def\thebibliography{\DeclareRobustCommand{\VAN}[3]{##3}\VANthebibliography}

\usepackage{graphicx}	
\usepackage{amsmath}	

\title[SPARCS Onboard Dynamic Exposure Control]{Onboard Dynamic Image Exposure Control for the \emph{Star-Planet Activity Research CubeSat (SPARCS)}}

\author[T. Ramiaramanantsoa et al.]{
Tahina Ramiaramanantsoa,$^{1}$\thanks{E-mail: tahina@asu.edu}
Judd D. Bowman,$^{1}$
Evgenya L. Shkolnik,$^{1}$
R. O. Parke Loyd,$^{1}$\newauthor
David R. Ardila,$^{2}$
April Jewell,$^{2}$
Travis Barman,$^{3}$
Christophe Basset,$^{2}$
Matthew Beasley,$^{4}$
Samuel Cheng,$^{2}$\newauthor
Johnathan Gamaunt,$^{1}$
Varoujan Gorjian,$^{2}$
John Hennessy,$^{2}$
Daniel Jacobs,$^{1}$
Logan Jensen,$^{1}$
Mary Knapp,$^{5}$\newauthor
Joe Llama,$^{6}$
Victoria Meadows,$^{7}$
Shouleh Nikzad,$^{2}$
Sarah Peacock,$^{8}$
Paul Scowen$^{1}$
and Mark R. Swain$^{2}$
\\
$^{1}$School of Earth and Space Exploration, Arizona State University, 781 E. Terrace Mall, Tempe, AZ 85287, USA\\
$^{2}$Jet Propulsion Laboratory, California Institute of Technology, 4800 Oak Grove Dr., MS 306-392, Pasadena, CA 91109, USA\\
$^{3}$University of Arizona, Lunar and Planetary Laboratory, 1629 E University Boulevard, Tucson, AZ 85721, USA\\
$^{4}$Southwest Research Inc., 1050 Walnut St \#300, Boulder, CO 80302, USA\\
$^{5}$Massachusetts Institute of Technology Haystack Observatory, 99 Millstone Road, Westford, MA 01886, USA\\
$^{6}$Lowell Observatory, 1400 W Mars Hill Rd, Flagstaff, AZ 86001, USA\\
$^{7}$Department of Astronomy, University of Washington, 3910 15th Ave NE, Seattle, WA 98195, USA\\
$^{8}$NASA Goddard Space Flight Center, Greenbelt, MD 20771, USA
}

\date{Accepted 2021 November 18. Received 2021 November 17; in original form 2021 October 22}

\pubyear{2021}

\begin{document}
\label{firstpage}
\pagerange{\pageref{firstpage}--\pageref{lastpage}}
\maketitle

\begin{abstract}
The \emph{Star-Planet Activity Research CubeSat} (\emph{SPARCS}) is a 6U CubeSat under development to monitor the flaring and chromospheric activity of M~dwarfs at near-ultraviolet (NUV) and far-ultraviolet (FUV) wavelengths. The spacecraft hosts two UV-optimized delta-doped charge-coupled devices fed by a 9-cm telescope and a dichroic beam splitter. A dedicated science payload processor performs near real-time onboard science image processing to dynamically change detector integration times and gains to reduce the occurrence of pixel saturation during strong M~dwarf flaring events and provide adequate flare light curve structure resolution while enabling the detection of low-amplitude rotational modulation.  The processor independently controls the NUV and FUV detectors.  For each detector, it derives control updates from the most recent completed exposure and applies them to the next exposure.  The detection of a flare event in the NUV channel resets the exposure in the FUV channel with new exposure parameters.  Implementation testing of the control algorithm using simulated light curves and full-frame images demonstrates a robust response to the quiescent and flaring levels expected for the stars to be monitored by the mission. The \emph{SPARCS} onboard autonomous exposure control algorithm is adaptable for operation in future point source-targeting space-based and ground-based observatories geared towards the monitoring of extreme transient astrophysics phenomena.
\end{abstract}




\section{Introduction}
\label{sec:intro}

High-cadence, high-quality, long-term monitoring of stars is best achieved from space-based observatories. They efficiently solve the issue of daytime and weather-related interruptions, achieve asymptotically diffraction-limited imaging, and provide access to wavelengths absorbed by the Earth's atmosphere.

\subsection{Photometric monitoring of stars done from a small satellite platform}
\label{subsec:intro_smallsats_for_stellar_astrophysics}

The past two decades have particularly seen a rise in successful small satellite platforms (mass $\leq$500~kg) dedicated to time-dependent stellar astrophysics.  These missions adopted reasonable trade-offs between cost, lifetime, lead time, complexity, duty cycle, and science objectives. The \emph{Microvariability and Oscillations of STars} \citep[\emph{MOST}:][]{2003PASP..115.1023W} mission pioneered the microsat regime ($10 - 100$~kg). The mission lasted much longer than its expected one-year lifetime. Over $10$ years of science operations, \emph{MOST} provided the astronomical community with new insights into the broadband optical light variability and internal structures of various types of stars across the Hertzsprung-Russell diagram, as well as exoplanets \citep[e.g.][]{2004PASP..116.1093R,2005ApJ...634L.109L,2005ApJ...623L.145W,2006ApJ...642..470A,2006ApJ...650.1111S,2008ApJ...689.1345R,2011ApJ...729...20Z,2012PASP..124..545H,2013ApJ...772L...2D,2014MNRAS.441..910R,2015MNRAS.446.4008E,2016ApJ...829L..31D}.  Then the \emph{BRIght Target Explorer} mission \citep[\emph{BRITE-Constellation:}][]{2014PASP..126..573W,2016PASP..128l5001P} pushed further into the nanosat regime ($1 - 10$~kg). Operating since early 2013, \emph{BRITE} has revealed new properties of the atmospheres and  winds of the brightest stars in the sky (V$\lesssim$8) through dual-band optical photometric monitoring \citep[e.g.][]{2016A&A...588A..55P,2017MNRAS.464.2249H,2017MNRAS.467.2494P,2018MNRAS.473.5532R,2018MNRAS.480..972R,2019MNRAS.490.5921R,2020NatAs...4..776A}. Recently, the \emph{Arcsecond Space Telescope Enabling Research In Astrophysics (ASTERIA)} has demonstrated the possibility to achieve the precision photometry needed to detect super-Earth transits from a CubeSat platform \citep{2020AJ....160...23K}.

These time-dependent stellar astrophysics SmallSat missions all probe stellar light variability in the optical domain. 
The \emph{Star-Planet Activity Research CubeSat} mission \citep[\emph{SPARCS}:][]{2018AAS...23122804S,2018arXiv180809954A} will spearhead the photometric monitoring of low-mass stars at ultraviolet (UV) wavelengths from a CubeSat platform. The 6U ($30$~cm~$\times$~$20$~cm~$\times$~$10$~cm) CubeSat is currently half-way in its development phase and is expected to be ready for launch in 2023. It is devoted to dual-band, high-cadence, time-resolved photometry of M~dwarfs to characterize their chromospheric and flaring activity in the UV. Such an endeavour is needed, not only in view of the increasing number of exoplanets discovered in the habitable zones of M~dwarfs \citep[e.g.][]{2016Natur.536..437A,2017Natur.542..456G,2019A&A...627A..49Z}, but also as M~dwarf frequent flaring events in the UV are theoretically found to affect the habitability and atmospheric loss of their planets \citep[e.g.][]{2010AsBio..10..751S,2015AsBio..15..119L}.  \emph{SPARCS} will provide empirical UV flaring statistics to inform these theoretical studies and allow future exoplanet observations to have an empirical UV context.

The \emph{Hubble Space Telescope (HST)} has been used to monitor a few M~dwarfs in the UV, leading to the detection of relatively strong M~dwarf UV flares peaking at $\sim$$500-14000$ times above quiescent level \citep[e.g.][]{2013ApJ...763..149F,2014ApJS..211....9L,2018ApJ...867...70L,2018ApJ...867...71L,2021ApJ...911L..25M}. The longest UV monitoring of an M~dwarf with \emph{HST} lasted $\sim$$30$~h. Yet, to establish accurate M~dwarf UV flare frequency distribution and prediction of the history and variability of their planets' UV exposure, longer UV monitoring of a larger sample of M~dwarfs is needed. Over an expected mission lifetime of $1$~yr, \emph{SPARCS} aims to probe the near-ultraviolet (NUV) and far-ultraviolet (FUV) light variability of $20$ young and old M~dwarfs on timescales ranging from minutes to weeks, by continuously monitoring each target between one and three stellar rotations ($5-30$)~d. The mission will measure flare color, energy, occurrence rate, and duration of quiescent and flare states for relatively active (young) and inactive (old) M~dwarfs. These time-dependent UV flux measurements are expected to further enable the construction of a new M~dwarf model atmosphere grid \citep{2019ApJ...886...77P,2020ApJ...895....5P}, during quiescent and flare states, and provide tighter contraints on the effects of M~dwarf UV radiation to the habitability and atmospheric loss of their planets.

\begin{figure*}
\includegraphics[width=17cm]{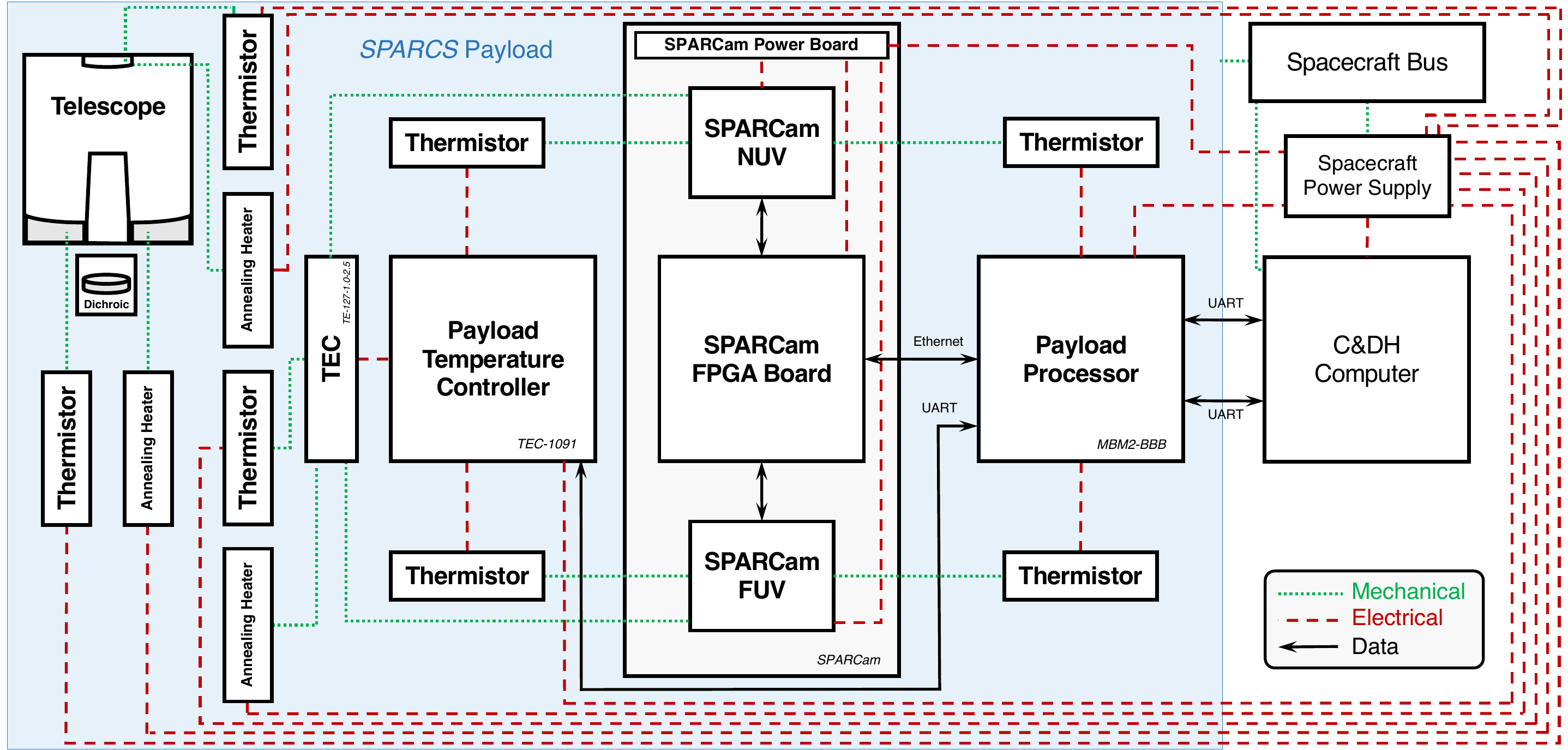}
\caption{\emph{SPARCS} payload architecture block diagram, laying out all the components of the payload, along with their interfaces between each other and to the relevant spacecraft elements that are external to the payload system.}
\label{fig:PL_Arch_C1}
\end{figure*}

\subsection{Dynamic image integration control onboard a space-based astronomical observatory}
\label{subsec:intro_dyn_exp_ctrl}

The \emph{SPARCS} science payload has its own processor board operating a Python-based custom payload software that performs command handling, detector thermal control regulation, as well as near real-time science image processing with autonomous dynamic exposure control to reduce the occurrence of pixel saturation from strong flaring events and allow for the detection of both low-amplitude rotational modulation and flare light curve structures.

The first recorded use of an automated exposure control of a charge-coupled device (CCD) onboard a spaceborne astronomical observatory was that of the \emph{Yohkoh}/SXT (Soft X-ray Telescope) solar observatory \citep{1991SoPh..136...37T}, which was implemented to mitigate pixel saturation when solar flares happen. The algorithm consisted in counting the number of pixels having values above a predefined upper limit value and below a predefined lower limit: if enough pixels (typically more than ten) exceeded the upper limit, the exposure time is decreased; if enough pixels (typically more than a hundred) were below the lower limit, the exposure time is increased. When overexposure or underexposure was detected, the selection of the subsequent exposure time invoked a database of predefined exposure times. Other space-based solar observatories such as \emph{TRACE} \citep[\emph{Transition Region and Coronal Explorer}:][]{1999SoPh..187..229H}, \emph{Hinode}/XRT \citep[X-ray Telescope:][]{2008SoPh..249..263K}, \emph{SDO}/AIA \citep[\emph{Solar Dynamics Observatory}/Atmospheric Imaging Assembly:][]{2012SoPh..275...17L} and \emph{IRIS} \citep[\emph{Interface Region Imaging Spectrograph}:][]{2014SoPh..289.2733D} adopted \emph{Yohkoh}/SXT's dynamic exposure control algorithm or a slight variant of it. So far, there is no record of autonomous dynamic exposure control operating onboard a space-based point source-targeting astrophysics observatory.

We hereby report on the development and performance of the \emph{SPARCS} onboard dynamic image exposure control functionality, which adopts an algorithm based on measurements of the maximum of the primary science target's point spread function (PSF) instead of a pixel count thresholding approach, and will be the first of its kind on a space-based time-dependent stellar astrophysics mission. In Section~\ref{sec:payload}, we describe the \emph{SPARCS} science payload. Section~\ref{sec:sciobs} is dedicated to the detailed description of the onboard dynamic image exposure control itself: the algorithm, the necessary onboard near real-time image processing, the photometry simulations used for testings, as well as testing results and pre-flight performances. Concluding remarks are provided in Section~\ref{sec:concl}.

\section{The SPARCS science payload}
\label{sec:payload}

\subsection{Science payload hardware}
\label{subsec:payload_hw}

The \emph{SPARCS} science payload fills a 3U volume ($30$~cm~$\times$~$10$~cm~$\times$~$10$~cm) in the 6U spacecraft. Figure~\ref{fig:PL_Arch_C1} illustrates the components of the payload along with their interfaces. Photon collection is performed by a $9$~cm f/6 Ritchey--Chr\'{e}tien telescope, after which a dichroic mirror splits the input beam into one that ultimately hits a NUV-optimized ($258-308$~nm) CCD and another one that falls on an FUV-optimized ($153-171$~nm) CCD. The \emph{SPARCS} camera (\mbox{SPARCam}) is the ensemble formed by the two sensor boards hosting the two CCDs, a Field Programmable Gate Array (FPGA) board, and a power board. \mbox{SPARCam} baselines the \mbox{CCD47-20} detector manufactured by Teledyne-e2v: a frame transfer, back-illuminated, \mbox{$1024$ $\times$ $1024$}, $13$-$\mu$m pixel format CCD with a thickness of $13$~$\mu$m. The camera uses Teledyne-e2v's Advanced Inverted Mode Operation (AIMO) version of the \mbox{CCD47-20}, which allows for a reduced dark signal. JPL's post-fabrication delta-doping process was used to tailor detector sensitivity for \emph{SPARCS} \citep{2018SPIE10709E..0CJ}. Additionally, an anti-reflection coating was applied to the NUV detector to improve quantum efficiency in the NUV band; this detector will be coupled with a stand-alone, commercial filter to block out-of-band signal, or ``red leak''. A metal-dielectric bandpass filter was integrated directly with the delta-doped FUV detector. The integrated filter optimizes transmission in the \emph{SPARCS} FUV band and mitigates red leak.

\begin{figure}
\includegraphics[width=\columnwidth]{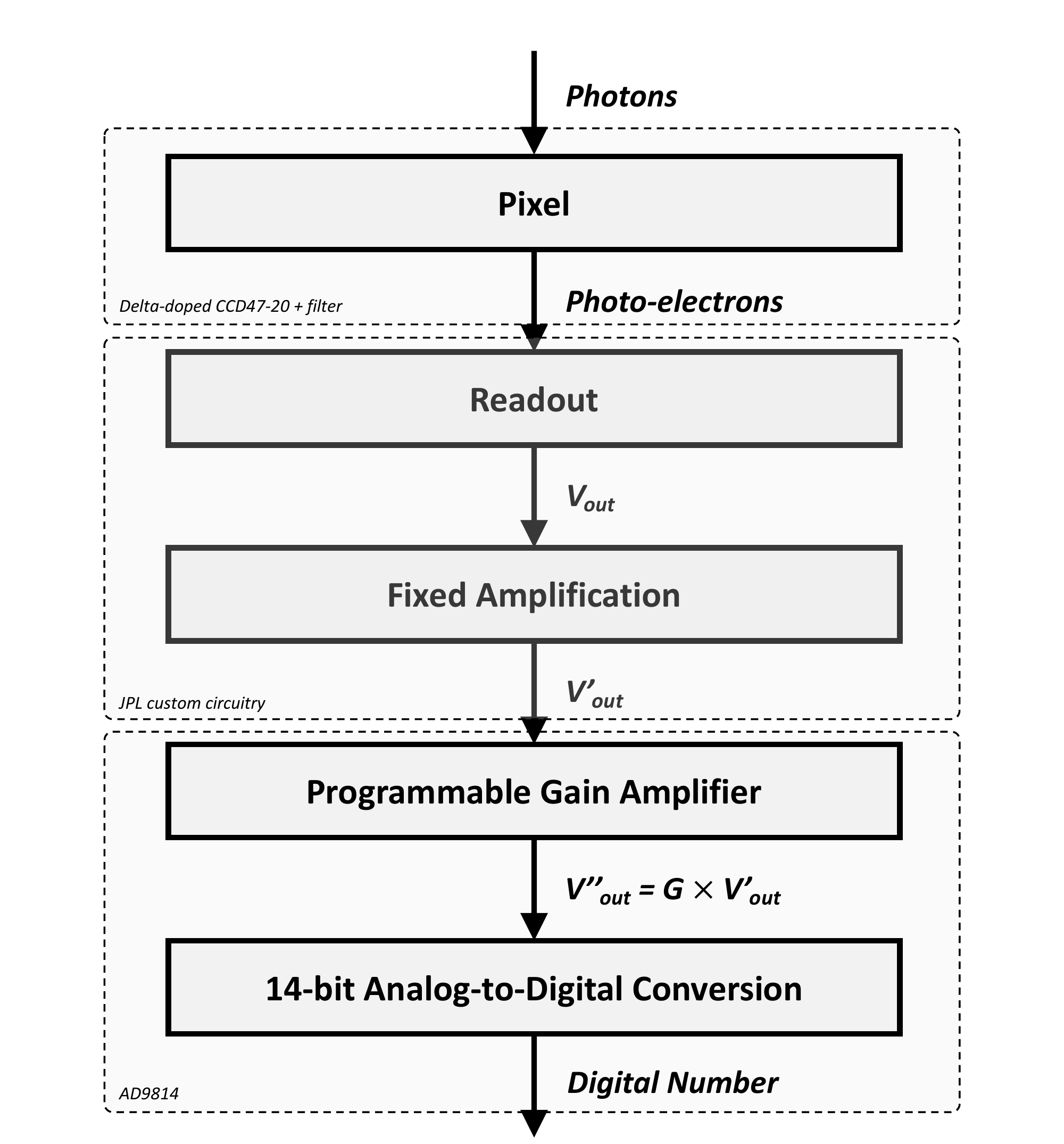}
\caption{SPARCam signal chain from photo-electron generation at the detector pixel level to digitization.}
\label{fig:sparcam_measurement_chain}
\end{figure}

Figure~\ref{fig:sparcam_measurement_chain} details the \mbox{SPARCam} signal chain from detector to digitization. Typical full well capacity is $100000$~electrons. Readout voltage signals get a fixed amplification to better match with the input voltage range of a \mbox{14-bit} analog-to-digital converter (ADC). The latter further amplifies the resulting voltage with a programmable analog gain $G$, then performs digitization. Under the assumption of ideal system parameter values, the system end-to-end gain $\kappa$ (electrons/ADU) and the adjustable analog gain follow the first-order analytical relation $\kappa\simeq6.7821/G$.

\begin{figure*}
\includegraphics[width=17cm]{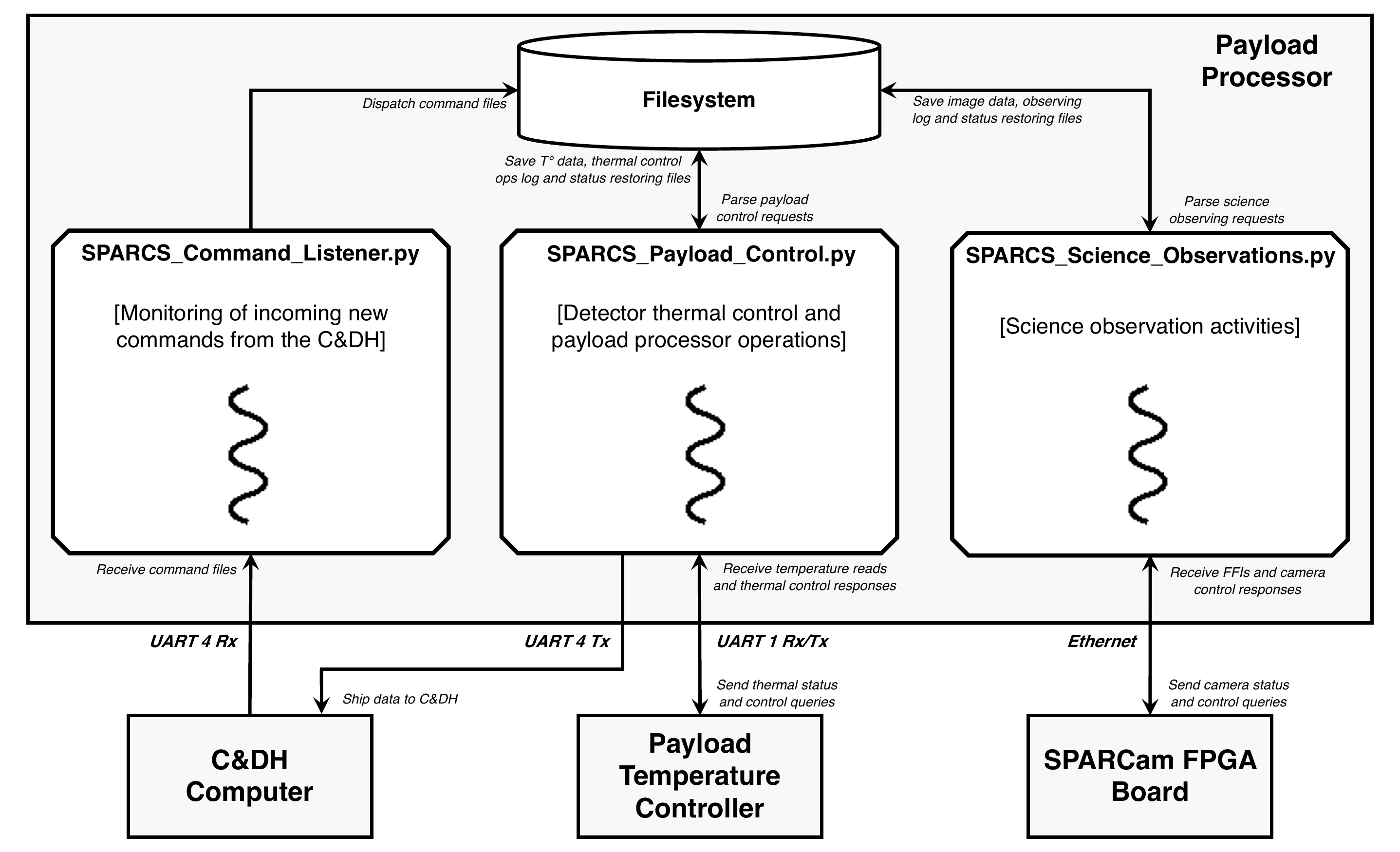}
\caption{High-level overview block diagram of the \emph{SPARCS} payload software, with highlights on the interactions between the software scripts and the payload processor directory system as well as the C\&DH computer, the payload temperature controller, and .}
\label{fig:PLP_SW_Config}
\end{figure*}

Full-frame images recorded by the CCDs are read out and buffered in a first-in first-out (FIFO) memory on the FPGA board, then transferred to a payload processor board via Ethernet for processing and subraster extraction. The payload processor also manages a temperature controller board that provides detector temperature measurements and regulation by means of a thermoelectric cooler (TEC). The payload processor board receives incoming commands from the spacecraft's command and data handling (C\&DH) computer and transfers payload data (raw science images, calibration images, time-tagged detector temperature measurements) to the C\&DH. Along with spacecraft telemetry data, payload data are ultimately transmitted from the C\&DH computer to the ground in S-band.

The \emph{SPARCS} payload processor board is a BeagleBone Black (BBB) with a Motherboard Module 2 (MBM2) protective breakout board manufactured by Pumpkin Space Systems. It has flight heritage in low-Earth orbit (LEO). The BBB adopts a 1-GHz ARM\textsuperscript{\tiny\textregistered} Cortex\textsuperscript{\tiny\textregistered}-A8 32‑bit processor, has $512$~MB Double Data Rate 3 (DDR3) random access memory, and is equipped with a $4$~GB $8$-bit embedded Multi-Media Controller (eMMC) flash storage and a microSD card slot. In addition to flight heritage, ground-based total ionizing dose (TID) tests of the BBB have demonstrated its survival up to a TID of $\sim$$170$~Gy, with the microSD card found to be the most susceptible component \citep{2015Kief}. Mission radiation analysis expects the \emph{SPARCS} payload processor to be exposed to a TID $\sim$$2.2$~Gy/yr in its planned Sun-synchronous LEO.

\subsection{Science payload software}
\label{subsec:payload_sw}

A KubOS\footnote{https://www.kubos.com/} Linux operating system runs on the \emph{SPARCS} payload processor.  The operating system supports Monit and is paired with U-boot to provide a fault-tolerant and recoverable environment. \emph{SPARCS} payload operation is performed through the execution of custom Python-based software.

The \emph{SPARCS} payload software is a Python 3.5.3-based package composed of three main scripts reliant on ten custom modules. As illustrated in Figure~\ref{fig:PLP_SW_Config}, the scripts run as three independent processes, with the first  monitoring a serial port for new commands from the C\&DH, the second dedicated to detector thermal control and payload processor maintenance operations, and the third managing science observations.  The three processes communicate through the filesystem, avoiding the use of a separate database and yielding persistent storage of commands and messages across reset events.

\section{The \emph{SPARCS} dynamic exposure control observing mode}
\label{sec:sciobs}

When the component of the \emph{SPARCS} payload processor software that manages science observations launches, it starts by restoring science observations to their latest saved state from a restoring file. Observing requests are queued and executed according to their scheduled execution dates. Following the execution of an observing request, a corresponding log file is created capturing the outcome of the commands and the science observation restoring file is updated.

The \emph{SPARCS} payload software has the capability to perform photometric monitoring in two modes.  The first uses simple fixed detector exposure times and gains. This could be selected to monitor fields that do not contain extreme flare stars such as M~dwarfs, e.g. during an extended mission. The second mode is the default mode of operation for the flare-monitoring campaigns in the \emph{SPARCS} nominal mission: dynamic detector exposure time and gain control, as described below. 

\subsection{Dynamic detector exposure time and gain control}
\label{subsec:sciobs_dec}

Previous monitoring of M~dwarf activity in the UV found that their UV flare amplitudes can reach \mbox{$\sim$$500-14000$} times their quiescent flux level \citep[][]{2018ApJ...867...70L,2018ApJ...867...71L,2021ApJ...911L..25M}. To account for such flare amplitudes, the detector exposure times have to be shortened whenever flares are detected in order to avoid pixel saturation as much as possible. This also leads to higher cadence in the flares, enabling more accurate determination of their energies \citep{2018ApJ...859...87Y}, while the longer out-of-flare exposures optimize S/N during quiescence for the study of rotational modulation related to chromospheric activity. For a required quiescent integration time, stacking of shorter sub-exposures is a less viable solution as it would yield lower S/N and thus require to monitor the target over more rotational cycles to better trace its rotational modulation variability. \emph{SPARCS} M~dwarf monitoring campaigns will always start with the detector gains set at their lowest possible values, and observations of flaring events will be first performed at reduced exposure times, in which case digital saturation could be reached even before the electronic well fills up \citep{2006hca..book.....H,2003hstc.conf..346M}. When the detector integration times reach their minimum allowable values, the detector gains are in turn increased to gain space in the ADC, until the electronic well saturates.

The \emph{SPARCS} onboard payload software is able to perform dynamic exposure control in both mono-band and dual-band modes of observations. In dual-band mode, as the \emph{SPARCS} NUV channel has higher sensitivity than the FUV channel, the former will always drive the dynamic exposure control: if the beginning of a flaring event is detected in the NUV, the (longer) concurrent exposure in the FUV channel is aborted and saved to disk, then new image captures are triggered in both channels with reduced exposure time, or increased gain if the exposure time is already at its minimum possible value. 

\begin{figure*}
\includegraphics[width=16.5cm]{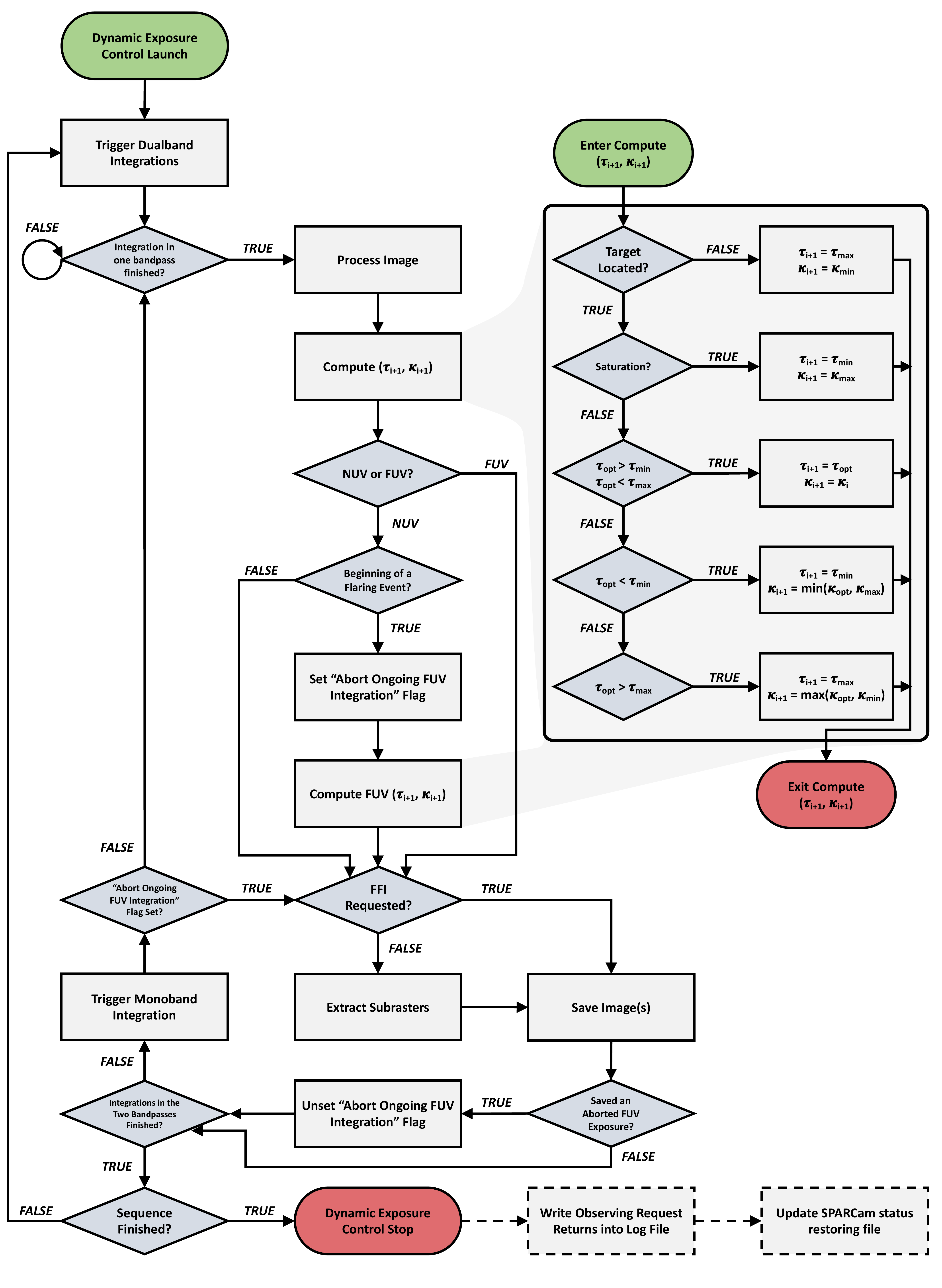}
\caption{Logic followed by the \emph{SPARCS} onboard dynamic science image integration control in dual-band mode of observation. The inset to the right of the diagram details the decision tree in the calculation of subsequent detector exposure times and gains.}
\label{fig:sciobs_dec_logic}
\end{figure*}

The optimal strategy for the dynamic exposure control is primarily dictated by the need to minimize the occurrence of pixel saturation in the primary science target's PSF, with minimal execution time and resource consumption. Hence, measuring the total target flux through standard aperture photometry \citep[e.g.][]{1992ASPC...23...90D} would be overkill and unhelpful as it does not capture information on single pixel saturation. Instead, measuring the maximum raw pixel value within the target's PSF is sufficient for tracking single pixel saturation. Figure~\ref{fig:sciobs_dec_logic} depicts the algorithm followed by the \emph{SPARCS} onboard dynamic exposure control in dual-band mode.

\begin{figure*}
\includegraphics[width=\textwidth]{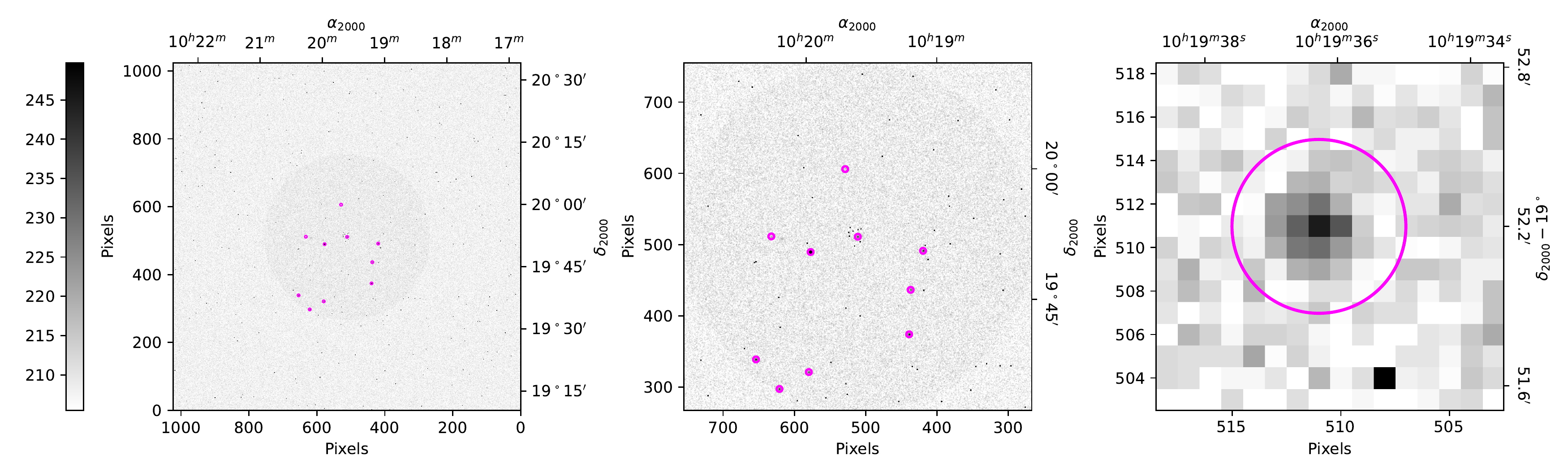}
\caption{\emph{Left:} Simulated \emph{SPARCS} raw $1$-min full frame image in the NUV channel for the field of view containing the M4.5V star AD Leo as primary science target located at the center of the image. Small circles pinpoint sources extracted by the payload processor software. \emph{Middle:} Zoom on a $40\arcmin\times40\arcmin$ window (i.e. the actual \emph{SPARCS} field-of-view) centered on the science target of interest. \emph{Right:} A $16$~pix $\times$ $16$~pix subraster containing the primary science target of interest, with the circle denoting the region in which the payload processor software probes the target PSF's maximum involved in the dynamic exposure control algorithm. The isolated black pixel towards the bottom of the subraster is a cosmic ray hit.}
\label{fig:PLP_SW_SciObs_FFI}
\end{figure*}

In the $i$-th raw image acquired within an integration time $\tau_i$ and at a detector gain $\kappa_i$ (electrons/ADU), the primary science target's PSF is distributed over a couple of pixels (Figure~\ref{fig:PLP_SW_SciObs_FFI}), with at least one pixel carrying the discretized PSF's global maximum $S_{{\rm max}, i}$. Thus, the bias-subtracted value of that maximum can be expressed as:
\begin{equation}
\widetilde{S}_{{\rm max}, i} = \frac{\tau_i}{\kappa_i}\left[\mathcal{F}_i\left(\Lambda_{e, i} + \beta_{e, i}\right) + \delta_{e, i} \right],
\label{eq:max_psf}
\end{equation}
where $\Lambda_{e, i}$ is the primary science target's count rate at that pixel (electrons/s/pix), $\beta_{e, i}$ the amount of sky background at that pixel (electrons/s/pix), $\delta_{e, i}$ the dark current rate (electrons/s/pix), and $\mathcal{F}_i$ the amount of photo response non-uniformity at that pixel. The \emph{SPARCS} dynamic exposure control algorithm considers a predefined setpoint value $\widetilde{S}_{\rm max, SP}$ for the bias-subtracted maximum of the primary target's PSF, and calculates the optimal subsequent exposure time to be the current one multiplied by the corrective factor needed to reach $\widetilde{S}_{\rm max, SP}$ based on the measured $\widetilde{S}_{{\rm max}, i}$ in the current image:
\begin{equation}
\tau_{\rm opt} = \frac{\widetilde{S}_{\rm max, SP}}{\widetilde{S}_{{\rm max}, i}}\times\tau_i.
\label{eq:optimal_texp}
\end{equation}

As long as $\tau_{\rm opt}$ remains within predefined allowable exposure time boundaries, the subsequent exposure time $\tau_{i+1}$ is set to $\tau_{\rm opt}$ and the detector gain remains unchanged. If the calculated optimal exposure time falls below its minimum allowable value $\tau_{\rm min}$, the subsequent exposure time is set to that minimum value and the detector gain is increased as follows:
\begin{equation}
\kappa_{\rm opt} = \frac{\tau_{\rm min}}{\tau_{\rm opt}}\times\kappa_i,
\label{eq:optimal_gain}
\end{equation}
until it reaches its maximum allowable value. That scheme is executed when pixel saturation does not occur in the primary science target's PSF. If the algorithm detects that a pixel has saturated, the subsequent exposure time is immediately set to the minimum possible value and the subsequent gain to its maximum allowable value. In total, this process is most analogous to the proportional component of a standard proportional-integral-derivative control loop.  The need to respond very rapidly at the start of flare events using only comparatively low-time resolution information reduces the usefulness of the integral and derivative components.  Hence, they are omitted from the \emph{SPARCS} dynamic exposure control logic.

The setpoint $\widetilde{S}_{\rm max, SP}$, along with exposure time and gain boundaries, are parts of a set of observing parameters that can be configured from the ground through observing commands. Among those parameters, $\widetilde{S}_{\rm max, SP}$ and the maximum allowable exposure time $\tau_{\rm max}$ are the ones that ultimately define how well-resolved the flares will be and what the shortest resolvable flares will be. For instance, if the NUV channel is configured such that its $\tau_{\rm max} = 10$~min and its $\widetilde{S}_{\rm max, SP}$ is set to a value that is achievable in $10$~min when the target star is quiescent, then most flares that last $\lesssim$$10$~min will not be resolved and, therefore, they will be more likely to cause pixel saturation. It is best to set $\tau_{\rm max}$ only just large enough to achieve the minimum desired signal-to-noise ratio (S/N) in quiescent phase at that exposure time, and set $\widetilde{S}_{\rm max, SP}$ to a moderately high value that is not achievable within $\tau_{\rm max}$ when the star is quiescent, but achievable during flare rising and decay phases. In that configuration, the exposure time will always max out at $\tau_{\rm max}$ during quiescent phase (in order to try to reach the unachievable $\widetilde{S}_{\rm max, SP}$), but will be variable and well-controlled within flaring events, which will also be well-resolved. Consequently, the condition that defines the detection of the beginning of a flaring event (decision box ``Beginning of a flaring event?'' in Figure~\ref{fig:sciobs_dec_logic}) and subsequently sets the flag to abort the ongoing FUV integration is: $\tau_{i} = \tau_{\rm max}$ \emph{and} $\tau_{i+1} \ne \tau_{\rm max}$.

For very low-S/N sources, care must be taken to provide margin for noise when setting $\tau_{\rm max}$ just above the minimum time needed to achieve the desired S/N in quiescent phase.  Otherwise, a slight apparent increase of flux in the target's PSF due to noise might cause the algorithm to reduce the subsequent exposure time and unnecessarily trigger an abort of the FUV exposure. There might be an oscillation between those two states until an actual flare occurs. This effect is most significant for very low-S/N stars that are barely observable in quiescent state with the observatory. An alternative observing strategy for these stars is discussed in Section~\ref{subsec:sciobs_dynamic_texp_g_testing}.

In addition to S/N considerations, this dynamic exposure control algorithm will also always yield a lag in the adjustment of the exposure parameters.  It is a feedback system with time delay: the appropriate exposure parameters can only be applied to the next exposure, whereas they should have been already adopted for the current exposure. It is therefore anticipated that a few cases of saturated frames might occur at the onset of very strong flaring events. We plan to investigate the potential of customized aperture photometry or even PSF fitting photometry in recovering accurate photometric measurements from these rare cases in ground-based data processing. Nevertheless, while it is practically impossible to achieve instantaneous dynamic control responses given the readout and transfer time of the detectors, one way to improve the accuracy of the algorithm would be to use the last exposure to predict the flux at the next integration based on a template for a typical flare light curve. Such a template would need to be well-constrained from observations to be useful. Also, given the diversity of UV flare temporal profiles \citep{2014ApJS..211....9L}, automating the prediction of the temporal shape of a flare from its first few moments will remain a formidable task and is not included in the dynamic control algorithm. Templates for M~dwarf NUV and FUV flare light curves, analogous to the optical flare shape template of the mid-M~dwarf GJ 1243 built from 6100 flares observed by the \emph{Kepler} mission from 11 months of monitoring of the star \citep{2014ApJ...797..122D}, might turn out to be among the by-products of the \emph{SPARCS} mission itself.

\subsection{Near real-time image processing}
\label{subsec:sciobs_improc}

The dynamic exposure control algorithm requires that each science image acquired subsequently receives some minimal image reduction (process box ``Process Image'' in Figure~\ref{fig:sciobs_dec_logic}). The reduction is needed to support source finding and involves bias frame subtraction, dark current frame subtraction, flat-fielding, bad pixel and cosmic ray cleaning.

The payload processor will be equipped with pre-flight bias frames and flat field frames acquired with detector gains ranging from $1.17$ to $6.78$. Similarly, pre-flight dark frames with the same gain values, but also acquired for different detector temperatures ranging from $-38\degr$C to $-33\degr$C with increments of $0.1\degr$C, will be pre-loaded on the payload processor. Those dark frames are acquired at the longest allowable \emph{SPARCS} detector integration time of $30$~min. For a given science image taken at a given detector gain, exposure time, and temperature, the bias frame with a gain closest to the image gain setting is selected and subtracted from the image. Then the dark frame with gain and temperature closest to those at which the image was acquired is selected, scaled to the actual exposure time of the image, and subtracted from the bias-corrected image. Finally, the flat field frame with gain closest to the gain setting of the image is chosen and divided out of the bias- and dark-corrected science image.

Pre-flight bad pixel masks marking the locations of hot, warm, and cold pixels on the two detectors will also be pre-loaded on the payload processor. The bad pixel masks will need to be regularly updated throughout the mission by taking snapshots of a region of the sky devoid of discernible sources. The \emph{SPARCS} payload software uses a simple thresholding scheme to identify cosmic rays in each exposure: it performs a $3$~pix $\times$ $3$~pix scan of a $32$~pix $\times$ $32$~pix region at the center of the image frame where the primary science target source is expected. Upon identification of a cosmic ray hit, a $3$~pix $\times$ $3$~pix median interpolation is performed. The choice of a restricted $32$~pix $\times$ $32$~pix region instead of the full frame image is to reduce execution time and resource consumption. The bad pixel and cosmic ray correction procedures are applied to the restricted $32$~pix $\times$ $32$~pix regions in both the bias/dark/flat-corrected science image and the bias-subtracted science image: the former is used for source finding, while the latter is invoked for dynamic exposure control.

Source localization is performed on the cleaned fully calibrated science image using a Python wrapper to the C-based package {\sc Source Extractor} \citep{1996A&AS..117..393B}. The primary target's expected location on the detector is compared to the list of source coordinates detected by {\sc Source Extractor}, and the coordinates that are closest to the expected coordinates are identified as the actual primary target centroid.

Once the primary science target's centroid is determined, the bias-subtracted maximum of the primary target's PSF $\widetilde{S}_{\rm max}$ is assessed within a circular aperture centered at the target's centroid in the cleaned bias-subtracted image. Although a circular aperture has to be placed around the primary source of interest to measure $\widetilde{S}_{\rm max}$, there is more freedom in the choice of the radius of that aperture than if we were to do standard aperture photometry, for which an aperture radius of one PSF full-width-at-half-maximum is usually found to optimize the signal-to-noise ratio \citep{2006hca..book.....H}. The \emph{SPARCS} dynamic exposure control algorithm's aperture radius can be configured from the ground as part of observing commands and defaults to $4$~pix (Figure~\ref{fig:PLP_SW_SciObs_FFI}). In that regard, as the plate scale is expected to be $\sim$$5$~arcsec/pix and the PSF anticipated to have a full width at half maximum of $\sim$$3$~pix, the spacecraft pointing stability (required to be less than $6$~arcsec [$1\sigma$] in $10$~min) does not affect the onboard dynamic exposure control logic.

Following the processing of an image and computation of the subsequent exposure time(s) and gain(s), either the raw full-frame image (FFI) or a set of \mbox{$16$~pix $\times$ $16$~pix} CCD subrasters containing the primary and some secondary targets extracted from the raw FFI is saved on the payload processor's disk to be transferred to the C\&DH computer and ultimately to the ground.

\subsection{Light curve and full-frame image simulations}
\label{subsec:sciobs_sparcsim_lite}

Testing of the \emph{SPARCS} onboard image processing and dynamic exposure control algorithm was performed using synthetic light curves and full-frame images in the two \emph{SPARCS} passbands.

Time-dependent flare flux measurements normalized to the Si~{\sc iv}~$\lambda\lambda 1394,1403$ doublet line flux were simulated following the method and assumptions described in \citet{2018ApJ...867...71L}. Normalization with respect to the Si~{\sc iv} line is adopted mostly because it is a strong line and the most sensitive to flares. Also, in that modeling approach, the temporal profile of an M~dwarf UV flare is roughly approximated by a sharp rising phase following a step function from quiescent level up to a peak amplitude, followed by a brief plateau, after which the flare amplitude decays exponentially. True M~dwarf UV flares will vary from this fiducial profile with more gradual rises, one or more sharp peaks, and variable rates of decay \citep[e.g.][]{2014ApJS..211....9L,2018ApJ...867...70L,2018ApJ...867...71L,2021ApJ...911L..25M}. The fiducial flare temporal profile was established as a simplified approximation to the observed UV flare on the M3.5 dwarf \mbox{GJ~876} \cite[see Figure 20 in][]{2018ApJ...867...71L}. First, we checked the response of the dynamic exposure control algorithm to the idealized flare temporal profile, as it offers the worst-case scenario in terms of flare rise steepness. Then we tested the response of the control algorithm to flare temporal profiles similar to that of the observed GJ~876 UV flare.

Once the time-resolved Si~{\sc iv}-normalized flare flux measurements $\phi(t)$ are established, the full stellar light curve is:
\begin{equation}
F(t) = \rho \phi(t) F_{{\rm Si~{\sc iv}, q}} + \psi(t) \int_{\lambda_{\rm min}}^{\lambda_{\rm max}} F_{{\rm \lambda, q}} Q(\lambda)\theta(\lambda)d\lambda, 
\label{eq:sim_flux}
\end{equation}
where $F(t)$ denotes the stellar flux (erg s$^{-1}$ cm$^{-2}$) at a given time $t$ in the bandpass of interest, $\psi(t)$ being variability owing to rotational modulation (see next paragraph), $F_{ {\rm \lambda, q}}$ the star's quiescent flux density in the bandpass, $Q(\lambda)$ the detector quantum efficiency, $\theta(\lambda)$ the filter transmittance, $F_{{\rm Si~{\sc iv}, q}}$ the star's quiescent Si~{\sc iv} flux, and $\rho$ represents the band-to-Si~{\sc iv} flux ratio ($5.094$ and $3.549$ for the \emph{SPARCS} NUV and FUV bandpasses, respectively). The band-to-Si~{\sc iv} flux ratio were computed based on the median energy budget of flares analyzed by \citet{2018ApJ...867...71L}. Implicit in our formulation is the assumption that the flare profiles are consistent between emission sources (e.g. UV continuum and  Si~{\sc iv}) and that the energy budget of flares does not vary between events, although there will be variations in reality. Then, for a given star, the Si~{\sc iv} flux is taken from previous studies that acquired UV spectra of M~dwarfs \citep[e.g.][]{2018ApJS..239...16F}, while the star's bandpass quiescent flux density $F_{ {\rm \lambda, q}}$ is estimated from its simulated quiescent spectrum covering the \emph{SPARCS} bandpasses \citep{2020ApJ...895....5P}.

\begin{figure*}
\includegraphics[width=\textwidth]{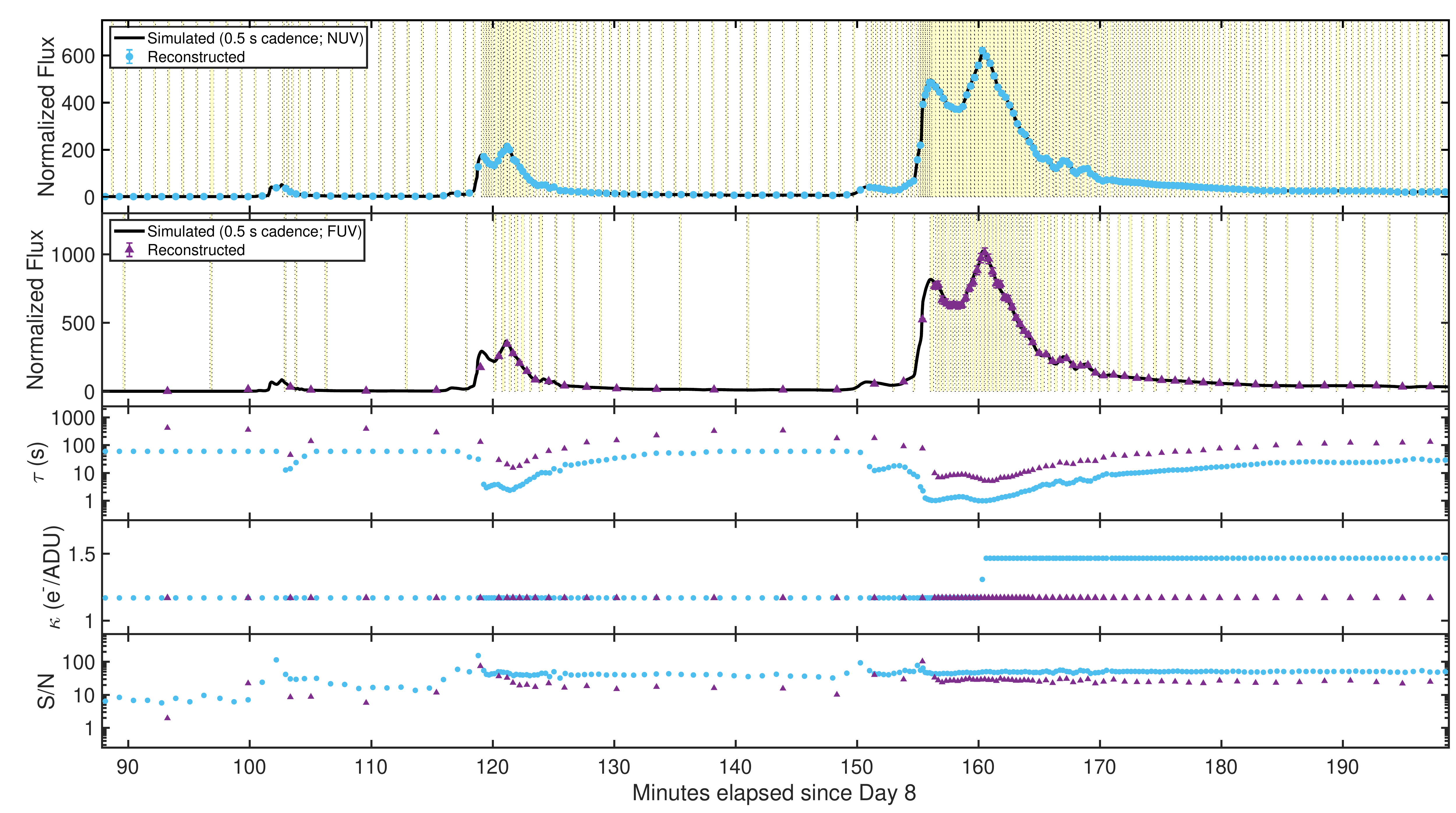}
\caption{Testing of the \emph{SPARCS} payload processor image processing and dynamic exposure control schemes. The top two panels show light curves of the M4 dwarf AD Leo (\emph{GALEX} NUV and FUV magnitudes: 15.8, 17.1) in the \emph{SPARCS} NUV and FUV passbands, respectively. In this case, when the star is in its quiescent phase, the exposure times reach the configured maximum $\tau_{\rm max}=1$~min in the NUV and $\tau_{\rm max}=7$~min in the FUV. Then the exposure times become shorter (and the gains higher when needed) upon flare detection. Shaded regions are times of no observations due to overheads. The three remaining bottom panels plot the evolution of the exposure time ($\tau$), the detector gain ($\kappa$), and the signal-to-noise ratio (S/N). By dynamically adjusting exposure times and gains, \emph{SPARCS} mitigates the occurence of detector pixel saturation during extreme flares and resolves flare light curve structures while optimizing S/N in quiescent states.}
\label{fig:PLP_SW_SciObs_LCs_Bright}
\end{figure*}

Given that previous UV observations of M~dwarfs reported rotational modulation with amplitudes up to $\sim$$25\%$ \citep{2017AJ....154...67M,2019A&A...629A..47D}, we also include rotational modulation in our simulated light curves. This is done by modulating the star's quiescent flux at the $25\%$ level through simulating rotational modulation arising from a mid-latitude bright chromospheric active region coming in and out of view as the star rotates, adopting the direct analytical spot modeling formalism introduced by \citet[][Equations 1 and 2 therein]{2003A&A...403.1135L}. Since there is currently limited statistical information on M~dwarf rotational modulation amplitude differences in the NUV and FUV, we chose them both to be $25\%$ in our light curve models. However, as described in Section~\ref{subsec:sciobs_dec}, the observing strategy can be tuned so that these relatively small variations do not affect exposure lengths during quiescent phases. The maximum exposure time $\tau_{\rm max}$ can be set to achieve a  desired S/N during the brightest portions of the rotational modulation ensuring each integration reaches the maximum exposure time across the entire slowly-varying, low-amplitude rotational modulation outside of flares.

\begin{figure*}
\includegraphics[width=\textwidth]{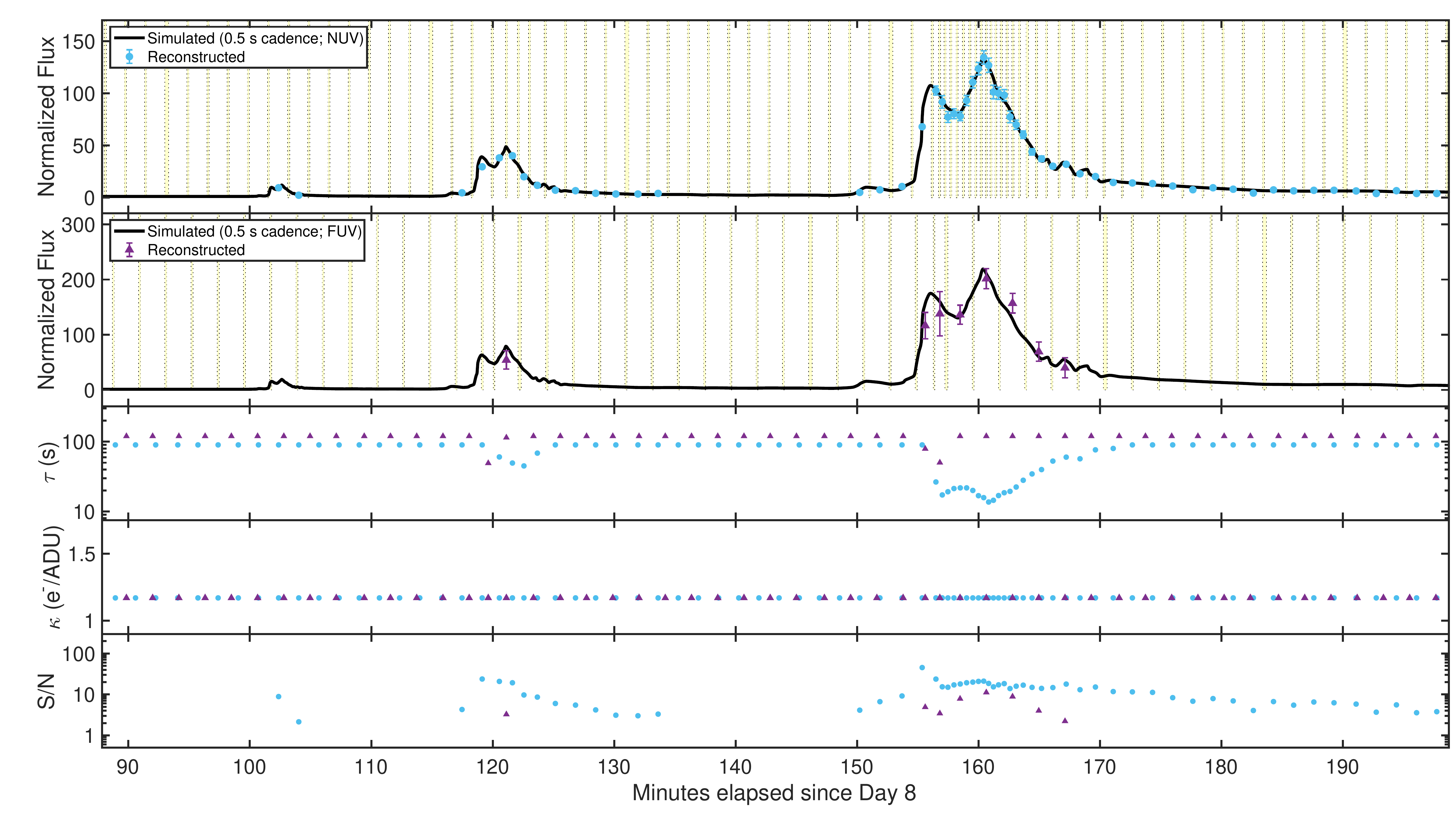}
\caption{Same as in Figure~\ref{fig:PLP_SW_SciObs_LCs_Bright} but for the fainter star GJ 832 (M2V; \emph{GALEX} NUV and FUV magnitudes: 18.4, 20.7). In this case, the star is not detected during its quiescent state when exposure times reach the configured maximum of $1.5$~min in the NUV and $2$~min in the FUV channel. Then the star is detected when it is flaring (see text).}
\label{fig:PLP_SW_SciObs_LCs_Faint}
\end{figure*}

For testing the image processing and dynamic exposure control, whenever an image capture is triggered, the image simulator is invoked with the requested exposure time $\tau$ and detector gain $\kappa$, along with a uniformly randomized detector temperature in the range $(-35\pm3)\degr$C.  The image simulator retrieves from the {\sc simbad} database all sources that are within a radius of $20\arcmin$ around the desired primary target and brighter than $20^{\rm th}$ magnitude in the \emph{GALEX (Galaxy Evolution Explorer)} NUV passband. Then the primary target count accumulated within the exposure integration time is evaluated as \citep[e.g.][]{2005PASP..117..421T}:
\begin{equation}
\Gamma =  \frac{\eta(1-\epsilon)\Omega\mathcal{A}\lambda}{\kappa hc} \int_{t_{0}}^{t_{0}+\tau} F(t) dt, 
\label{eq:sim_Nph}
\end{equation}
with $\mathcal{A}$ the effective \emph{SPARCS} telescope diameter ($8.27$~cm), $\Omega$ the throughput of the optical telescope assembly ($67\%$), $\eta$ the dichroic throughput ($90\%$ in the NUV, $60\%$ in the FUV), and $\epsilon$ the amount of red leak ($2\%$ in the NUV, $55\%$ in the FUV)\footnote{The adopted amount of red leak in the FUV channel is a representative worst-case scenario (inactive star) value. Typical best-case scenario (active star) value would be $2\%$. In any case, all quoted instrument parameter values are either current best estimates or conservative values, pending full characterization once the instrument is fully assembled. A description of the \emph{SPARCS} instrument model will be the subject of a forthcoming publication (Ardila et al., in prep.).}. A simulated full-frame image populated by Gaussian PSFs at the locations of the sources is then generated, with added sky background ($1\times10^{-17}$~erg/s/cm$^2$/\AA/arcsec$^2$ in the NUV, $2\times10^{-18}$~erg/s/cm$^2$/\AA/arcsec$^2$ in the FUV) and dark current noise following Poisson distributions, as well as a readout noise following a Gaussian distribution ($\sigma = 4$~electrons/pix) and an offset of $200$~ADU.  Cosmic ray hits are added to the simulated image adopting the \emph{HST}'s Wide-Field Camera~3 cosmic ray event rate of $1.2$~event/s/cm$^2$ and an average deposition per event of $2300$~electrons \citep{GreenOlszewski2020}.  Figure~\ref{fig:PLP_SW_SciObs_FFI} shows an example simulated \emph{SPARCS} science image in the NUV channel.

\subsection{Testing results and pre-flight performances}
\label{subsec:sciobs_dynamic_texp_g_testing}

\begin{figure*}
\includegraphics[width=\textwidth]{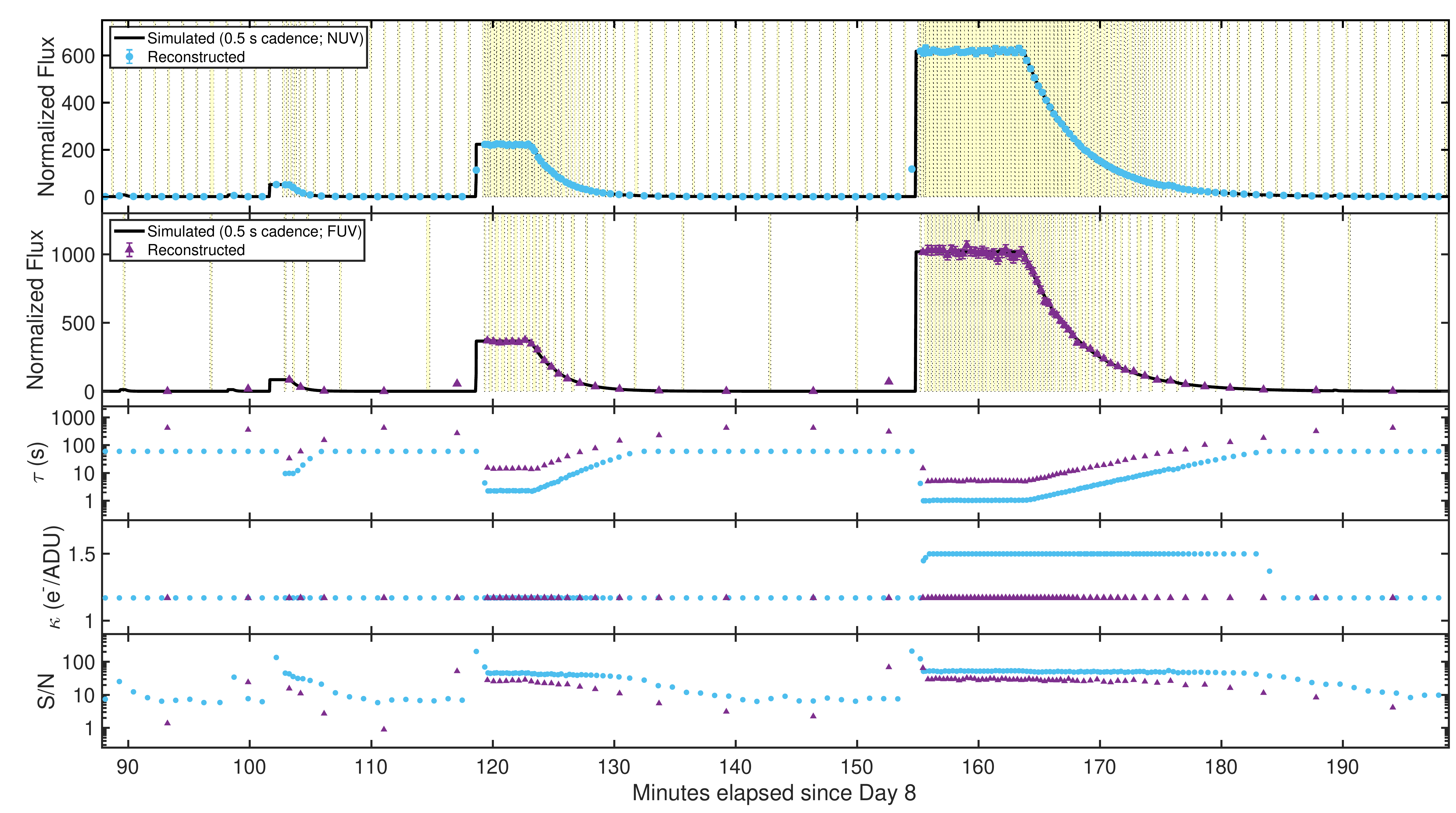}
\caption{Same as in Figure~\ref{fig:PLP_SW_SciObs_LCs_Bright} but for the worst-case scenario of a step-wise flare rise offered by the fiducial flare temporal profile.}
\label{fig:PLP_SW_SciObs_LCs_Bright_Fiducial}
\end{figure*}

Figures~\ref{fig:PLP_SW_SciObs_LCs_Bright}~and~\ref{fig:PLP_SW_SciObs_LCs_Faint} show excerpts of the implementation testing of the \emph{SPARCS} onboard image processing procedure and dynamic exposure control algorithm using the photometry simulator. The plots show two cases, a bright target (AD Leo)  and a faint target (GJ 832), respectively. In the first two panels of these plots, the continuous black lines are simulated noise-free light curves with $0.5$~s cadence (Equation~\ref{eq:sim_flux}), while the larger points that come with $1\sigma$ uncertainties (blue filled circles in the NUV, purple triangles for the FUV) are light curves reconstructed from simple circular aperture photometry on the simulated images generated by the photometry simulator. The $1\sigma$ uncertainties take into account the target source's shot noise, read noise, dark current noise, sky background noise, red leak noise, and telescope pointing jitter noise. The response of the dynamic exposure control algorithm to the actual observed UV flare temporal profile is satisfactory. The control algorithm has also demonstrated good response to the worst-case scenario of a step-wise flare rise (Figure~\ref{fig:PLP_SW_SciObs_LCs_Bright_Fiducial}).

The faint star case illustrated in Figure~\ref{fig:PLP_SW_SciObs_LCs_Faint} is noteworthy as it shows the expected behaviour when the maximum exposure time $\tau_{\rm max}$ is set such that the target star would not be detected in individual exposures by either the NUV and FUV channels during quiescence, and would only be detected when it is flaring. This approach can be adopted if the required quiescent exposure time would be too long to allow resolving the relatively short-duration flaring events (as short as the ones in the models). A set of longer integrations could be acquired at the beginning of the observing campaign in order to capture the target source's median quiescent fluxes in the NUV and FUV. Then the rest of the campaign could be performed with $\tau_{\rm max}$ short enough to capture flaring events only, as depicted in Figure~\ref{fig:PLP_SW_SciObs_LCs_Faint}. This observing strategy may not allow tracking of the rotational modulation component of the stellar variability, but does enable the study of the flaring events on targets that are at the edge of being too faint to be observable in quiescent phase with the observatory.

A delay of at least $\sim$$9.7$~s always occurs between the end of an image acquisition at the detector level and its final storage on the payload processor due to image transfer from the imaging part of the detector to the storage part, image readout into the FIFO memory on the FPGA board, image transfer from \mbox{SPARCam} to the payload processor, image assembling, image reduction, source extraction, exposure control, and image writing onto the payload processor's eMMC flash storage. Longer overhead may happen in either channel at times when an image integration was finished but its transfer and processing have to wait because the software is in the middle of the processing of another image acquired in the other channel. Such a situation is most likely to occur during strong flaring events, when the exposure times are so short that the system is constrained to operate at reduced duty cycle. Also, since the payload processor has to read images from a FIFO memory on the SPARCam FPGA board, it is not possible to transfer a specific subraster in the middle of the full-frame image (where the primary M dwarf target is located) from the camera to the payload processor without first transferring at least about the bottom half part of the full-frame image. However, further optimization of the software and image processing steps may reduce the overall time overhead.

\section{Conclusion}
\label{sec:concl}

The NASA-funded \emph{SPARCS} mission is a stellar astrophysics CubeSat observatory for the investigation of the chromoshperic and flaring activity of M~dwarfs in the UV through long time baseline, high-cadence photometric monitoring using a 9-cm telescope coupled with a dual-band NUV/FUV CCD-based camera. A custom, fully Python-based payload software manages \emph{SPARCS} onboard payload operations and features a dynamic science image exposure control mode of observation that is primarily driven by the need to mitigate the occurence of pixel saturation in the primary science target's PSF during flaring events, as well as the need to both track low-amplitude rotational modulation variability and resolve flare structures. The \emph{SPARCS} onboard dynamic exposure control adopts an algorithm that autonomously decides on the optimal detector exposure times and gains based on the need to reach a given setpoint value for the bias-subtracted maximum of the primary science target's PSF.

The control algorithm was tested using synthetic light curves in the \emph{SPARCS} NUV and FUV passbands containing both flaring events and rotational modulation, and invoking a full-frame image simulator that accounts for read noise, dark current, sky background, and random cosmic ray hits. The tests revealed good response to an actual observed M~dwarf UV flare temporal profile, and even to the worst-case scenario of step-wise flare rises. Based on this testing, it is expected that the mission will best resolve strong M~dwarf UV flaring events such as the $\sim$$14000\times$ UV flare that has been recorded on Proxima Cen \citep{2021ApJ...911L..25M}.

\emph{SPARCS} will be the first space-based stellar astrophysics observatory that we know of to adopt an autonomous onboard dynamic science image integration control. Once the mission launches and records a reasonable number of M~dwarf UV flares that can be used to create a representative M~dwarf UV flare temporal profile, the \emph{SPARCS} dynamic exposure control algorithm could be updated to adopt a more predictive behavior, potentially involving machine learning or dynamic time warping approaches. Also, the control algorithm has a level of genericness that allows for a relatively easy adaptability for operation onboard other point source-targeting space-based observatories -- or even ground-based observatories -- dedicated to the study of extreme transient astrophysics phenomena using CCD-based imaging systems.

\section*{Acknowledgements}

The \emph{SPARCS} team thanks the reviewer of this article for their insightful constructive comments. The \emph{SPARCS} team gratefully acknowledges support from the National Aeronautics and Space Administration Astrophysics Research and Analysis (APRA) program (NNH16ZDA001N-APRA; 80NSSC18K0545). A portion of the research was carried out at the Jet Propulsion Laboratory, California Institute of Technology, under a contract with the National Aeronautics and Space Administration (80NM0018D0004).

\section*{Data Availability}

The data underlying this article will be shared on reasonable request to the corresponding author.



\bibliographystyle{mnras}
\bibliography{SPARCS_Onboard_Dynamic_Exposure_Control} 

%


\bsp	
\label{lastpage}
\end{document}